\documentclass[aps,prc,twocolumn,nofootinbib,amsmath,showpacs,
superscriptaddress,floatfix]{revtex4-1}
\usepackage{graphicx,epsfig,epstopdf,longtable}
\usepackage{mathptmx}
\usepackage{wasysym}
\usepackage{color}
\usepackage[flushleft]{threeparttable}

\newcommand{\beqy}{\begin{eqnarray}}
\newcommand{\eeqy}{\end{eqnarray}}
\newcommand{\bmlet}{\begin{subequations}}
\newcommand{\emlet}{\end{subequations}}
\newcounter{saveeqn}

\def\gsimeq{\,\,\raise0.14em\hbox{$>$}\kern-0.76em\lower0.28em\hbox  
{$\sim$}\,\,}  
\def\lsimeq{\,\,\raise0.14em\hbox{$<$}\kern-0.76em\lower0.28em\hbox  
{$\sim$}\,\,}  

\begin{document}

\title{Statistical properties of $^{243}$Pu, and $^{242}$Pu(n,$\gamma$)
cross section calculation}
\author{T.A.~Laplace}
\email{lapthi@berkeley.edu}
\affiliation{Lawrence Livermore National Laboratory, Livermore, CA 94551, USA}
\affiliation{Department of Nuclear Engineering, University of California, 
Berkeley, CA 94720, USA}
\author{F.~Zeiser}
\affiliation{Department of Physics, University of Oslo, N-0316 Oslo, Norway}
\author{M.~Guttormsen}
\affiliation{Department of Physics, University of Oslo, N-0316 Oslo, Norway}
\author{A.C.~Larsen}
\affiliation{Department of Physics, University of Oslo, N-0316 Oslo, Norway}
\author{D.L.~Bleuel}
\affiliation{Lawrence Livermore National Laboratory, Livermore, CA 94551, USA}
\author{L.A.~Bernstein}
\affiliation{Lawrence Livermore National Laboratory, Livermore, CA 94551, USA}
\affiliation{Department of Nuclear Engineering, University of California,
 Berkeley, CA 94720, USA}
\affiliation{Lawrence Berkeley National Laboratory, Berkeley, CA 94720, USA}
\author{B.L.~Goldblum}
\affiliation{Department of Nuclear Engineering, University of California, 
Berkeley, CA 94720, USA}
\author{S.~Siem}
\affiliation{Department of Physics, University of Oslo, N-0316 Oslo, Norway}
\author{F.L.~Bello~Garotte}
\affiliation{Department of Physics, University of Oslo, N-0316 Oslo, Norway}
\author{J.A.~Brown}
\affiliation{Department of Nuclear Engineering, University of California, 
Berkeley, CA 94720, USA}
\author{L. Crespo~Campo}
\affiliation{Department of Physics, University of Oslo, N-0316 Oslo, Norway}
\author{T.K.~Eriksen}
\affiliation{Department of Physics, University of Oslo, N-0316 Oslo, Norway}
\author{F.~Giacoppo}
\affiliation{Department of Physics, University of Oslo, N-0316 Oslo, Norway}
\author{A.~G{\"o}rgen}
\affiliation{Department of Physics, University of Oslo, N-0316 Oslo, Norway}
\author{K.~Hady\'{n}ska-Kl\c{e}k}
\affiliation{Department of Physics, University of Oslo, N-0316 Oslo, Norway}
\author{R.A.~Henderson}
\affiliation{Lawrence Livermore National Laboratory, Livermore, CA 94551, USA}
\author{M.~Klintefjord}
\affiliation{Department of Physics, University of Oslo, N-0316 Oslo, Norway}
\author{M.~Lebois}
\affiliation{Institut de Physique Nucl{\'e}aire d'Orsay, B{\^a}t. 100, 15 rue 
G. Clemenceau, 91406 Orsay Cedex, France}
\author{T.~Renstr{\o}m}
\affiliation{Department of Physics, University of Oslo, N-0316 Oslo, Norway}
\author{S.J.~Rose}
\affiliation{Department of Physics, University of Oslo, N-0316 Oslo, Norway}
\author{E.~Sahin}
\affiliation{Department of Physics, University of Oslo, N-0316 Oslo, Norway}
\author{T.G.~Tornyi}
\affiliation{Department of Physics, University of Oslo, N-0316 Oslo, Norway}
\author{G.M.~Tveten}
\affiliation{Department of Physics, University of Oslo, N-0316 Oslo, Norway}
\author{A.~Voinov}
\affiliation{Department of Physics and Astronomy, Ohio University, Athens, 
Ohio 45701, USA}
\author{M.~Wiedeking}
\affiliation{iThemba LABS, P.O. Box 722, Somerset West 7129, South Africa}
\author{J.N.~Wilson}
\affiliation{Institut de Physique Nucl{\'e}aire d'Orsay, B{\^a}t. 100, 15 rue 
G. Clemenceau, 91406 Orsay Cedex, France}
\author{W.~Younes}
\affiliation{Lawrence Livermore National Laboratory, Livermore, CA 94551, USA}

\date{\today}

\begin{abstract}
The level density and $\gamma$-ray strength function ($\gamma$SF) of $^{243}$Pu
have been measured in the quasi-continuum using the Oslo method. Excited states
in $^{243}$Pu were populated using the $^{242}$Pu$(d,p)$ reaction. The level
density closely follows the constant-temperature level density formula for
excitation energies above the pairing gap. The $\gamma$SF displays a
double-humped resonance at low energy as also seen in previous investigations
of actinide isotopes. The structure is interpreted as the scissors resonance
and has a centroid of $\omega_\mathrm{SR}=2.42(5)$~MeV and a total strength of
$B_\mathrm{SR}=10.1(15)$~$\mu_N^2$, which is in excellent agreement with
sum-rule estimates. The measured level density and $\gamma$SF were used to
calculate the $^{242}$Pu$(n, \gamma)$ cross section in a neutron energy range
for which there were previously no measured data. 
\end{abstract}


\maketitle

\section{Introduction}
\label{sec:int}
Neutron capture cross sections are required for the accurate modeling of
advanced nuclear energy systems and nucleosynthesis in neutron-rich
astrophysical environments. Unfortunately, it is often difficult to
accurately measure (n,$\gamma$) cross sections for short-lived ``minor''
actinides over the energy range of greatest relevance to nuclear energy and
astrophysical applications. In these cases an alternative approach is to
measure the properties of excited nuclear states, including the nuclear level
density and $\gamma$-ray strength function ($\gamma$SF), and use these as
inputs for calculations of neutron capture reaction rates using statistical
model codes. In this paper we are concentrating on the properties of the 
n+$^{242}$Pu compound system.

With a half life of 0.37 million years, $^{242}$Pu is the second longest-lived
isotope of Pu after $^{244}$Pu. Though its radioactivity is not one of the
largest contributors to nuclear waste decay heat, $^{242}$Pu is fissionable by
fast neutrons and can be recycled in fast reactors. With the increase of the
fuel cycle length and the development of fast reactors aimed at reducing
radioactive waste comes the need for reliable cross sections for a fast neutron
spectrum~\cite{aliberti2006,chadwick2011}. It is extremely important to be able
to accurately predict cross sections where measured data are insufficient or
nonexistent.

Furthermore, improvements in the modeled reaction rates could also improve
predictions of actinide abundances on Earth~\cite{ar07}. Actinides are
synthesized in extreme stellar environments uniquely by the rapid neutron
capture process~\cite{ar07}. For accurate predictions, reactions rates for
actinides with high neutron excess are the most crucial. However, such actinide
isotopes with extreme neutron-to-proton ratios are out of reach experimentally,
and will remain so for many years to come. Thus, it is imperative to obtain a
better understanding of the fundamental nuclear properties in this mass
region, to provide stringent test on nuclear models invoked to calculate these
reaction rates~\cite{ar07}.

Measurements of the statistical properties of the nucleus are important for
nuclear cross section calculations in the framework of the statistical model.
The nuclear level density and $\gamma$SF can be extracted using the Oslo
method~\cite{Schiller00,Lars11}. This method has been successfully applied
recently in the actinide region to the $^{231-233}$Th, $^{232,233}$Pa,
$^{237-239}$U and $^{238}$Np
isotopes~\cite{guttormsen2012,nld2013,gsf2014, tornyi14}. Thus far, the level
densities of all measured actinides follow closely the constant-temperature
level density formula. These heavy and well deformed systems also show a low
energy orbital $M1$ scissors resonance (SR).
The main purpose of the present work is to extract the level density and
$\gamma$SF in $^{243}$Pu. Comparing the measured to an estimated $\gamma$SF,
the SR is extracted and interpreted as an enhancement of the $\gamma$SF.
Hauser-Feshbach calculations of the neutron capture cross section using the
measured level density and $\gamma$SF as inputs are compared with evaluated
nuclear databases.

In Sec.~II the experimental procedure is described. Sec.~III discusses the
extraction and normalization of the level density and $\gamma$SF.
In Sec.~IV, the experimental SR is presented and theoretical sum rules are
briefly introduced. Extracted resonance parameters are compared to previous
results and sum-rules estimates. In Sec.~V, the measured level density and
$\gamma$SF are used as inputs to Hauser-Feshbach calculations in order to
estimate the $^{242}$Pu$(n,\gamma)$ cross sections. Conclusions are drawn
in Sec.~VI.

 \begin{figure*}[!t]
 \begin{center}
 \includegraphics[clip,width=2\columnwidth]{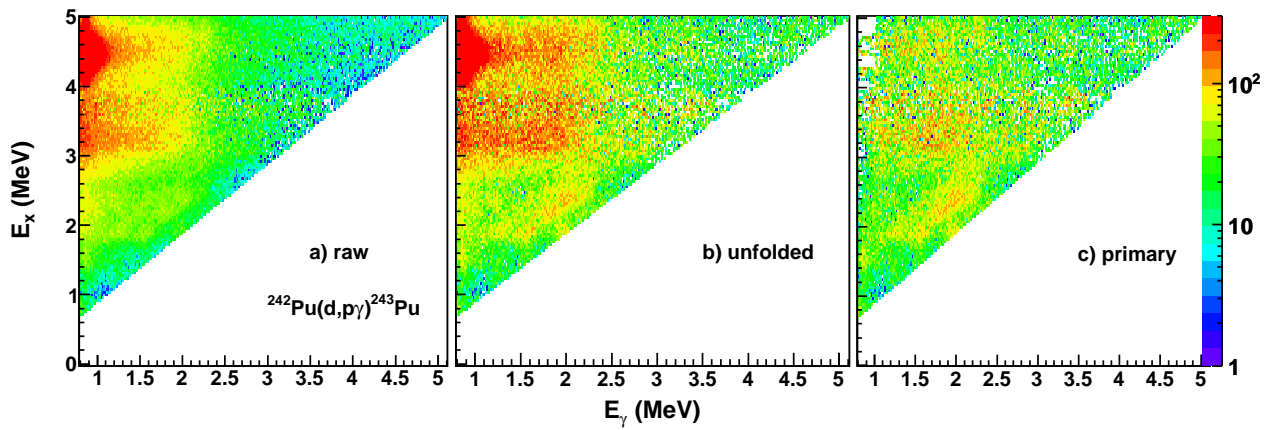}
 \caption{(Color online) Initial excitation energy $E_\mathrm{x}$ versus
 $\gamma$-ray energy $E_{\gamma}$ from particle-$\gamma$ coincidences recorded
 from the $^{242}$Pu$(d,p\gamma)^{243}$Pu reaction. The raw $\gamma$-ray
 spectra (a) are first unfolded (b) by the NaI response function. The primary
 or first-generation $\gamma$-ray spectra (c) are extracted as function of
 excitation energy $E_\mathrm{x}$.}
 \label{fig:matrix}
 \end{center}
 \end{figure*}
\section{Experimental methods}
\label{sec:exp}

The experiment was conducted using the MC-35 Scanditronix cyclotron at the
Oslo Cyclotron Laboratory (OCL). The 0.4 mg/cm$^2$ $^{242}$Pu on a
Be-backing target was bombarded with a 12~MeV deuteron beam with a beam
current of $\approx$~1~nA. Prior to electrodeposition, the Pu material was
cleaned from decay products and other impurities using an anion-exchange resin
column procedure~\cite{Hen11}. The purified product was electroplated onto a
thin Be foil (1.9 mg/cm$^2$ thickness) from a small aliquot of dilute nitric
acid placed into a large volume of isopropanol.  The resulting target was
dried, baked at $500^\circ$C in a muffle furnace, then glued to the target
frame. 

Particle-$\gamma$ coincidences were measured using the SiRi particle telescope
and the CACTUS $\gamma$-detector system~\cite{siri,CACTUS}. The SiRi particle
telescope is composed of eight segmented Si particle telescopes, which in this
experiment were placed at backward angles of $\theta = 126^\circ$ to
$140^\circ$ relative to the beam axis, drastically reducing the contribution
from elastically scattered deuterons. 
The resulting spin distribution of the nucleus is more representative of the
compound nuclei formed in higher-energy (e.g., non-thermal) (n,$x$) reactions.
Each telescope is comprised of a $\Delta$E and E Si detector with thicknesses
of $130$~$\mu$m and $1550$~$\mu$m, respectively.
The CACTUS array consists of 26 collimated
$5^{\prime\prime} \times 5^{\prime\prime}$ NaI(Tl) detectors surrounding the
target and particle telescopes, and with a total efficiency of $14.1$(2)$\%$
at $E_{\gamma} = 1.33$~MeV. 

The particle telescopes were used to generate the master gate signal and the
start signal for the Time-to-Digital-Converters (TDC). The NaI detectors were
used as individual TDC stops. Thus prompt particle-$\gamma$ coincidences with
background subtraction were sorted event by event. The proton events were
extracted by setting proper two-dimensional gates on the $\Delta$E-E matrices.
Using the measured proton energies deposited in the telescopes and the
reaction Q value, the initial excitation energy $E_\mathrm{x}$ in the residual
$^{243}$Pu nucleus was calculated. To avoid contamination from $\gamma$-rays
emitted by other reaction channels, only excitation energies below the fission
barrier ($B_f\approx 6$~MeV~\cite{RIPL3}) and the neutron separation energy
($S_n=5.034$~MeV~\cite{ENDF243Pu}) were considered.
 
The recorded particle-$\gamma$ coincident events were sorted into a
two-dimensional matrix as shown in Fig.~\ref{fig:matrix}~(a).
Using the known response function of the CACTUS array, the raw data were
unfolded to correct for the NaI response functions and efficiency, and regain
the full-energy peaks for each 40~keV excitation energy bin~\cite{Gutt96}.
The unfolded matrix is shown in Fig.~\ref{fig:matrix}~(b). A peak at
$E_\gamma$=870~keV from O contamination was subtracted from the matrix.

The Oslo method was used to extract the first generation (primary)
$\gamma$-rays from the total $\gamma$-ray cascade~\cite{Gutt87}. The main
assumption in the technique is that $\gamma$ decay from a given excitation
energy is independent on how the nucleus was excited (e.g., directly via
$(d,p)$ reactions or from $\gamma$ decay from a higher-lying level). The
first-generation $\gamma$-ray matrix $P(E_\mathrm{x},E_{\gamma})$ is shown
in Fig.~\ref{fig:matrix}~(c).

 \begin{figure}[h]
 \begin{center}
 \includegraphics[clip,width=\columnwidth]{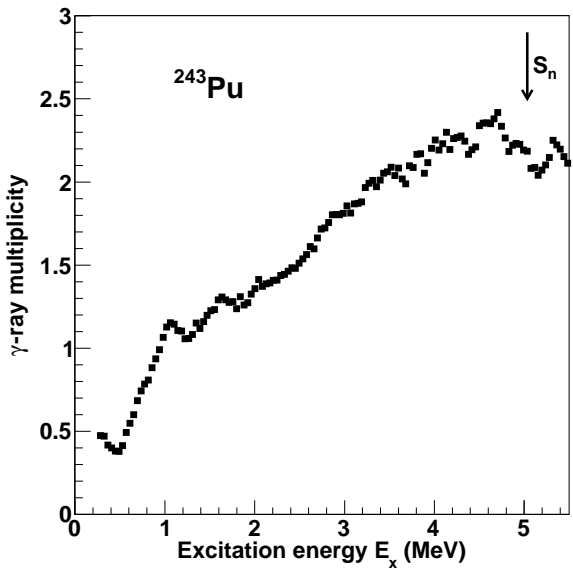}
 \caption{Gamma-ray multiplicity as function of excitation energy
 $E_\mathrm{x}$ in $^{243}$Pu. Only $\gamma$-rays with $E_{\gamma} > 0.4$~MeV
 are taken into account.}
 \label{fig:mult}
 \end{center}
 \end{figure}

Figure~\ref{fig:mult} shows the average $\gamma$-ray multiplicity
$\langle M_{\gamma}(E_\mathrm{x})\rangle$ for $E_{\gamma} > 0.4$~MeV as
function of initial excitation energy $E_\mathrm{x}$ given by:
\begin{equation}
\langle M_{\gamma}(E_\mathrm{x})\rangle=
\frac{E_\mathrm{x}}{\langle E_{\gamma}(E_\mathrm{x})\rangle},
\end{equation}
where the average $\gamma$-ray energy $\langle E_{\gamma}(E_\mathrm{x})\rangle$
is calculated from the unfolded $\gamma$ matrix at a fixed excitation energy
$E_\mathrm{x}$.

Below $E_\mathrm{x}=2$~MeV, the multiplicity fluctuates indicating a
non-statistical behavior of the decay process while approaching the ground
state.
Above $E_\mathrm{x}=4.5-5$~MeV, the multiplicity fluctuates due to the opening
of the fission and neutron emission channels. To apply the Oslo method, only
the $E_\mathrm{x}=2.6-4.3$~MeV region of the first generation matrix of
Fig.~\ref{fig:matrix}~(c) is used. 

Fermi's golden rule~\cite{Fermi1950} states that the decay probability can be
factorized into a transition matrix element between the initial and final state
and the density of final states. According to Brink's hypothesis~\cite{brink},
the transmission coefficient ${\cal {T}}$, which plays the role of the
transition matrix element in Fermi's Golden rule, is independent of the
excitation energy.
The first generation matrix $P(E_\mathrm{x},E_{\gamma})$ can be factorized as
follows:
\begin{equation}
P(E_\mathrm{x}, E_{\gamma}) \propto   
{\cal{T}}(E_{\gamma}) \rho (E_\mathrm{x} -E_{\gamma}),
\label{eqn:3}
\end{equation}
where $\rho (E_\mathrm{x} -E_{\gamma})$ is the level density at the 
excitation energy after the primary $\gamma$-ray has been emitted in
the cascades.
Simultaneous extraction of the level density and the $\gamma$-ray
transmission coefficient is achieved using an iterative procedure to the
first generation matrix $P(E_\mathrm{x},E_{\gamma})$~\cite{Schiller00}.
It has been shown~\cite{Schiller00} that if one solution for the
multiplicative functions $\rho$ and ${\cal{T}}$ is known, one may
construct an infinite number of transformations $\tilde{\rho}$ and
$\tilde{{\mathcal{T}}}$, which give identical fits to the
$P(E_\mathrm{x},E_{\gamma})$ matrix by:
\begin{eqnarray}
\tilde{\rho}(E_\mathrm{x}-E_\gamma)&=&A\exp[\alpha(E_\mathrm{x}
-E_\gamma)]\,\rho(E_\mathrm{x}-E_\gamma),
\label{eq:array1}\\
\tilde{{\mathcal{T}}}(E_\gamma)&=&B\exp(
\alpha E_\gamma){\mathcal{T}} (E_\gamma),
\label{eq:array2}
\end{eqnarray}
where the parameters $A$, $\alpha$ and $B$ cannot be determined by the
fitting procedure. Their determination is discussed in the next section.

 \begin{figure}[t]
 \begin{center}
\includegraphics[clip,width=\columnwidth]{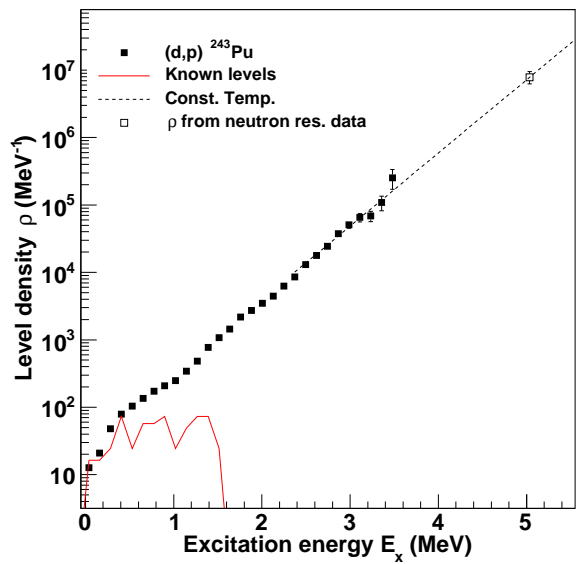}
 \caption{(Color online) Level density for $^{243}$Pu. At low excitation energy
 $E_\mathrm{x}$, the experimental data are normalized to the level density of
 known discrete levels (red solid line). At the neutron separation energy
 $S_n=5.038$ MeV, the normalization is done using the level density extracted
 from known neutron resonance spacings $D_0$. The connection between
 $\rho(S_n)$ (the upper right data point) and the experimental data 
 is made with a constant-temperature formula with $T_{\rm CT}= 0.40(1)$ MeV.}
 \label{fig:rhotot}
 \end{center}
 \end{figure}
%
    \begin{table*}[]
    \caption{Parameters used to extract level density and 
    $\gamma$SF (see text).}
    \begin{threeparttable}
    \centering
    \begin{tabular}{cccp{0.8cm}p{0.8cm}|cc|c}
    \hline
    \hline
    $S_n$ & $a$ & $E_1$   &  $\sigma(S_n)$&  $D_0$   &   $\rho(S_n)$ 
    &  $\rho(S_n)_{\rm red}$    &$\langle \Gamma_{\gamma}(S_n)\rangle$  \\
    (MeV)&(MeV$^{-1})$ & (MeV)&      & (eV)   &(10$^6$MeV$^{-1}$)
    &(10$^6$MeV$^{-1}$)& (meV)       \\
    \hline
    5.034& 25.82\tnote{a}      &-0.45\tnote{a}  & 8.15\tnote{a}  
    & 17(1)\tnote{b}  &    7.87(163)      &    3.94   &   22(1)\tnote{b} \\
    \hline
    \hline
    \end{tabular}
    \begin{tablenotes}
    \item[a] Estimated from systematics~\cite{egidy2}.
    \item[b] Ref.~\cite{mughabghab2006}.
    \end{tablenotes}
    \end{threeparttable}
    \label{tab:parameters}
    \end{table*}

\section{Normalization of the level density and $\gamma$SF}
The parameters $A$ and $\alpha$ of Eq.~(\ref{eq:array1}) are needed to obtain
a normalized level density. They can be determined by matching the data points
at low energy to known discrete levels~\cite{ENSDF} and estimating the level
density at the neutron separation energy $S_n$ from neutron-resonance spacing
data using the formula~\cite{Schiller00}:
\begin{equation}
\rho(S_n)=\frac{2\sigma^2}{D_0}\cdot \frac{1}{(I_t+1)
\exp\left[-(I_t+1)^2/2\sigma^2\right]+I_t\exp\left[-I_t^2/2\sigma^2\right]}
\end{equation}  
where $I_t$ is the spin of the target nucleus, $D_0$ the neutron resonance
spacing, and $\sigma$ the spin-cutoff parameter. The following spin
distribution is assumed~\cite{GC} in the produced nucleus:
\begin{equation}
g(E_\mathrm{x}=S_n,I) \simeq \frac{2I+1}{2\sigma^2}
\exp\left[-(I+1/2)^2/2\sigma^2\right],
\label{eq:spindist}
\end{equation}
where $I$ is the spin in the resulting nucleus. The spin-cutoff parameter
$\sigma$ was determined from the global systematic study of level-density
parameters by von Egidy and Bucurescu,
using a rigid-body moment of inertia approach~\cite{egidy2}:
\begin{equation}
\sigma^2 = 0.0146 A^{5/3} \frac{1+\sqrt{1+4aU}}{2a},
\label{eq:8}
\end{equation}
where $A$ is the mass number, $a$ is the level density parameter,
$U=E_\mathrm{x}-E_1$ is the intrinsic excitation energy, and $E_1$ is the
back-shift parameter. The $a$ and $E_1$ parameters are obtained from global
systematics.  
Table~\ref{tab:parameters} lists the parameters used to estimate $\sigma$.
From the deduced $\sigma$ value and $D_0$ at $S_n$ the level density $\rho$ is
determined.
Several values for $D_0$ were reported in the literature: RIPL-3 
($D_0=13.5(15)$~eV)~\cite{RIPL3},
Mughabghab ($D_0=17(1)$~eV)~\cite{mughabghab2006}, Young and Reeder 
($D_0=16.5$~eV)~\cite{Young70}, and Rich, \textit{et al.}, using the ESTIMA 
code ($D_0=16.8(5)$~eV)~\cite{Rich09}. The RIPL-3~\cite{RIPL3} value is 
inconsistent with other works. Normalization of the level density and the 
$\gamma$SF was
performed using the different values and using $D_0=17(1)$~eV provided a  
more consistent result with a measurement
on a $^{240}$Pu target made at the same facility~\cite{FabioPu240}.
Thus, to
obtain the level density at $S_n$ given in Table~\ref{tab:parameters}, the
$D_0$ parameter is taken from Ref.~\cite{mughabghab2006}. In order to perform
the normalization at $S_n$, the constant-temperature formula~\cite{CT} is used
for the interpolation of our experimental data points and the level density
at $S_n$:
\begin{equation}
\rho_\mathrm{CT}(E_\mathrm{x})=\frac{1}{T_\mathrm{CT}} \exp 
\frac{E_\mathrm{x} - E_0}{T_\mathrm{CT}}.
\label{eq:CT}
\end{equation}
The slope of the level density is given by $T_\mathrm{CT}=0.40~(1)$~MeV
and the shift in excitation energy by $E_0=-0.95~(16)$~MeV.
Figure \ref{fig:rhotot} shows the level density normalized at low and high
excitation energies. The level density follows closely the constant-temperature
formula with $\ln \rho \propto E_\mathrm{x}/T_{\rm CT}$ between
$E_\mathrm{x}\approx2$~MeV and $E_\mathrm{x}\approx3$~MeV as observed for other
actinide nuclei~\cite{nld2013,tornyi14}. The Fermi-gas model does not describe
properly the data as described in Ref.~\cite{Gutt15}.

The standard procedure to normalize the level density and $\gamma$SF is
problematic when a $(d,p)$ entrance channel is employed in actinides to form
the compound nucleus. The spin distribution in the compound nucleus populated
using the $(d,p)$ is not as broad as for other reactions such as
($^3$He,$\alpha$) which can bring in more angular momentum. 
As the slope, $\alpha$, of the transmission coefficient is the same for the
level density in Eqs.~(\ref{eq:array1}) and (\ref{eq:array2}), a reduced
level density $\rho_{\mathrm{red}}$ corresponding to the level density for the
spins populated in the $(d,p)$ reaction was used to obtain the correct slope
of $\mathcal{T}$. The reduced level density thus is extracted by assuming a
lower value of $\rho$ at $S_n$. This effect has been demonstrated in simulated
data using DICEBOX on the case of $^{163}$Dy~\cite{Lars11}, where a restriction
on the spin of the initial levels was made
($1/2  \leq I_{\mathrm{initial}} \leq 13/2$). To obtain a correct slope of the
transmission coefficient, and thus the $\gamma$SF extracted from the simulated
data (see Figs.19-21 in Ref.~\cite{Lars11}), the level density had to be
normalized not to the full level density, but to the reduced level density for
spins within the range $1/2  \leq I_{\mathrm{final}} \leq 15/2$ (one primary
transition of dipole type accounts for $I_{\mathrm{final}} = 15/2)$.

Cumulating large uncertainties in the total $\rho(S_n)$ and the unknown actual
spin distribution brought into the nuclear system by the specific $(d,p)$
reaction, the extracted slope of ${\cal{T}}$ becomes rather uncertain. Those
complications encountered using the $(d,p)$ reaction on actinides make the
standard normalization procedure of the Oslo method~\cite{Schiller00,voin1}
to determine the $\alpha$ parameter for the transmission coefficient in
Eq.~(\ref{eq:array2}) quite uncertain. 

The determination of the parameter $B$ of Eq.~(\ref{eq:array2}) gives the
absolute normalization of ${\cal{T}}$. The average total radiative width
$\langle \Gamma_{\gamma} \rangle$ at $S_n$, assuming that the $\gamma$-decay
in the continuum is dominated by dipole transitions, is used here for
normalization. The width is related to the transmission coefficient
${\cal{T}}$ by~\cite{ko90}: 
\begin{eqnarray}
\langle\Gamma_\gamma\rangle=\frac{1}{2\pi\rho(S_n, I, \pi)} 
\sum_{I_f}&&\int_0^{S_n}{\mathrm{d}}E_{\gamma} B{\mathcal{T}}(E_{\gamma})
\nonumber\\
&&\times \rho(S_n-E_{\gamma}, I_f),
\label{eq:norm}
\end{eqnarray}
where $I$ and $\pi$ are the initial spin and parity at $S_n$ respectively.
The summation and integration is performed over all final levels with spin
$I_f$ that are accessible by $E1$ or $M1$ transitions with energy $E_{\gamma}$.
The determination of $B$ using Eq.~(\ref{eq:norm}) is influenced by systematic
errors because the integral depends on both the level density
$\rho(E_\mathrm{x})$ and the spin-cutoff parameter $\sigma(E_\mathrm{x})$;
the latter quantity is not well constrained experimentally for all excitation
energies. Given these complications, we have compared the $\gamma$SF with the
extrapolation of known photonuclear reaction ata, in addition to determining
the $B$ parameter using Eq.~(\ref{eq:norm}) to reproduce the
experimentally-determined $\gamma$-width
$\langle\Gamma_\gamma\rangle$~\cite{mughabghab2006} listed in
Table~\ref{tab:parameters}. Hence, by making use of an independent experimental
constraint, we reduce the systematic uncertainties in the absolute
normalization of the $\gamma$SF.

 \begin{figure}[t]
 \begin{center}
 \includegraphics[clip,width=\columnwidth]{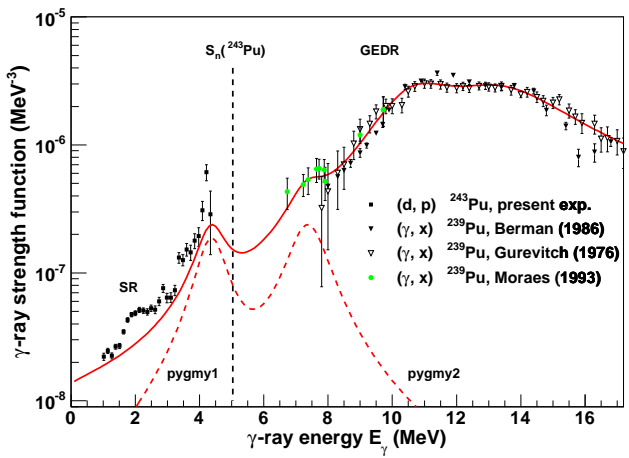}
 \caption{(Color online) Experimental $\gamma$SF from the present
 $(d,p\gamma)^{243}$Pu experiment (black filled squares) and ($\gamma$, x)
 data (black filled triangles, empty triangles and green dots) taken
 respectively from Berman {\em et al.}~\cite{berman1986},
 Gurevitch {\em et al.}~\cite{gurevitch1976} and
 Moraes {\em et al.}~\cite{Moraes1989}. The red curve represents the
 estimated underlying $E1$ part of the $\gamma$SF. The structure in the
 present work at E$_\gamma=1-3.5$~MeV is interpreted as the Scissors
 Resonance.}
 \label{fig:GDR}
 \end{center}
 \end{figure}
\begin{table*}[]
\caption{Resonance parameters used for the $\gamma$SF extrapolation.}
\begin{tabular}{ccc|ccc|c|ccc|ccc}
\hline
\hline
$\omega_{E1,1}$&$\sigma_{E1,1}$&$\Gamma_{E1,1}$&$\omega_{E1,2}$
&$\sigma_{E1,2}$&$\Gamma_{E1,2}$&$T_f$&$\omega_{\rm pyg1}$
&$\sigma_{\rm pyg1}$&$\Gamma_{\rm pyg1}$&$\omega_{\rm pyg2}$
&$\sigma_{\rm pyg2}$&$\Gamma_{\rm pyg2}$ \\
 (MeV)  &      (mb)     &      (MeV)    &   (MeV)  &     (mb)      
 &    (MeV)      &(MeV)&    (MeV)    &      (mb)       
  &          (MeV)    &(MeV)  &     (mb)    &  (MeV)    \\ \hline
11.1    &      290      &       3.2     &     14.2 &      340     
 &     5.5       & 0.40(1) &     4.4(1)    &      9(3)         
  &        1.0(2)        & 7.4(3) &     20(6)    &    1.3(3)   \\
\hline
\end{tabular}
\\
\label{tab:GDR_param}
\end{table*}

Given these complications, a different procedure is used, comparing the
$\gamma$SF with the extrapolation of known photonuclear reaction data.
The strength function is related to the transmission coefficient
${\cal {T}}(E_{\gamma})$ by~\cite{RIPL3}:
\begin{equation}
f (E_{\gamma}) \approx \frac{1}{2\pi} 
\frac{ {\mathcal{T}}(E_{\gamma})}{ E_{\gamma}^3}.
\label{eq:fT}
\end{equation}
These data are compared with the strength function derived from
the cross section $\sigma$ of photo-nuclear reactions by~\cite{RIPL3}:
\begin{equation}
f (E_{\gamma}) =\frac{1}{3\pi^2 \hbar^2c^2} 
\frac{\sigma(E_{\gamma})}{ E_{\gamma}},
\label{eq:fT2}
\end{equation}
where the 
factor $1/3\pi^2\hbar^2c^2 = 8.6737\times10^{-8}$~mb$^{-1}$MeV$^{-2}$.

In Fig.~\ref{fig:GDR} the $\gamma$SF derived from $^{239}$Pu($\gamma$, x)
cross sections by Berman~{\em et al.}~\cite{berman1986},
Gurevitch~{\em et al.}~\cite{gurevitch1976} and
Moraes~{\em et al.}~\cite{Moraes1989} are shown. There are no measured
$\gamma$SF data for other Pu isotopes. It is assumed that the E1 strength
does not vary much from $^{239}$Pu to $^{243}$Pu, as seen for similar mass
U isotopes~\cite{gsf2014}, and supported by the classical Thomas-Reiche-Kuhn
sum rule for E1 strength~\cite{thomas,reiche,kuhn}. 

The data from the present work cover $\gamma$ energies up to 4.3~MeV while
for the ($\gamma$, x) data, the lowest energy point is 6.7~MeV. Some
interpolation is needed to link the different data sets. The GEDR displays a
double-humped feature  common to all well-deformed nuclei that is fitted with
two enhanced generalized Lorentzians (EGLO) as defined in RIPL~\cite{RIPL3},
but with a constant-temperature parameter of the final states $T_f$, in
accordance with the Brink hypothesis. To take into account the steep rise
of our $\gamma$SF data from E$_\gamma=3-4$~MeV, a resonance is postulated
at around 4.4~MeV (labeled pygmy1 in Fig.~\ref{fig:GDR}). Due to the
absence of data between 4.3~MeV and 6.7~MeV, the parameters of the
resonance are highly uncertain.
In addition, the ($\gamma$, x) data~\cite{Moraes1989} reveal a knee at
around 7.5~MeV indicating an additional resonance-like structure
(labeled pygmy2 in Fig.~\ref{fig:GDR}). The two pygmy resonances
are described by standard Lorentzians:
\begin{equation}
f_{\rm pyg}=\frac{1}{3\pi^2\hbar^2c^2}\frac{\sigma_{\rm pyg}
\Gamma_{\rm pyg}^2E_{\gamma}}
{(E_{\gamma}^2 - \omega_{\rm pyg}^2)^2+ \Gamma_{\rm pyg}^2E_{\gamma}^2},
\end{equation}
where $\sigma_{\rm pyg}$, $\Gamma_{\rm pyg}$, $\omega_{\rm pyg}$ are the
strength, width, and energy centroid of the pygmy resonance, respectively. 

The red curve in Fig.~\ref{fig:GDR} is the sum of the GEDR and the two pygmy
resonances. It serves as a ``base line" of the $\gamma$SF with no additional
strength from other resonances. The parameters for the GEDR and the two pygmy
resonances are given in Table~\ref{tab:GDR_param}. The measured $\gamma$SF is
normalized to this underlying $E1$ strength. To match the slope of the observed
$\gamma$SF with the GEDR low-energy tail, the level density at $S_n$ was
reduced from 7.87 to 3.94~million levels per MeV thereby varying the $\alpha$
parameter from Eq.~(\ref{eq:array1}). Calculations of the spin population
using the Empire code~\cite{empire} suggest a reduced level density
$\rho_{\mathrm{red}}=0.34(4)\times\rho$ for $^{243}$Pu, $^{233}$Th and
$^{238}$Np. which were all produced using a 12~MeV deuteron beam. Because
the experimental level densities of $^{232}$Th, and $^{237}$Np were reduced
by a factor of $\approx 1/2$~\cite{gsf2014,tornyi14}, the same reduction
factor was used.

\section{The scissors resonance and sum rules}
\begin{figure}[t]
 \begin{center}
\includegraphics[clip,width=\columnwidth]{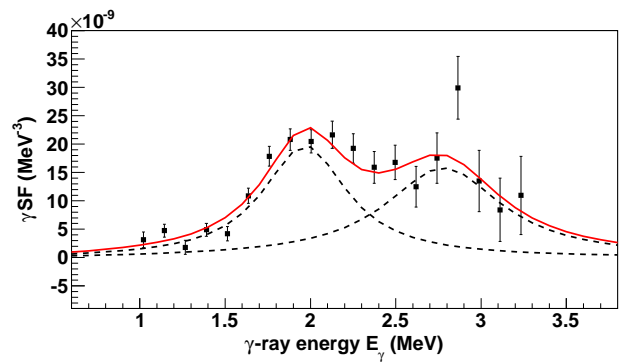}
 \caption{(Color online) The extracted $\gamma$SF for the scissors
 resonance in the quasi-continuum of $^{243}$Pu.}
 \label{fig:pygmy}
 \end{center}
 \end{figure}

Figure~\ref{fig:pygmy} shows a $\gamma$SF measured above the expected
$\gamma$SF base line (red curve in Fig.~\ref{fig:GDR}). The extra strength
between E$_\gamma=1$ and 3.5~MeV is interpreted as the ``scissors'' resonance
(SR). A similar structure has been previously observed in the $^{231-233}$Th,
$^{232,233}$Pa, $^{237-239}$U and $^{238}$Np
isotopes~\cite{guttormsen2012,nld2013,gsf2014,tornyi14}. Even though the
parameters of the resonance postulated at 4.5~MeV are rather uncertain, the SR
is extracted by subtracting a smoothly varying background under the $\gamma$SF
which is mainly composed of the temperature-dependent low energy tail of the
GEDR, as described in the previous section.
The SR has been fitted with two Lorentzians. The resonance centroid $\omega_i$,
cross section $\sigma_i$, and width $\Gamma_i$ for the lower ($i=1$) and upper
($i=2$) resonances are listed in Table~\ref{tab:strengths}, as well as the
total strength and average energy centroid.

 \begin{flushright}
 
 \end{flushright}

The separation between the two components, $\Delta\omega_{\rm SR}=0.81(6)$~MeV,
is similar to what has been observed for Th, Pa and U~\cite{gsf2014}
($\Delta\omega_{\rm SR}=0.89(15)$~MeV) and higher than the $^{238}$Np
observation, $\Delta\omega_{\rm SR}=0.53(6)$~MeV~\cite{tornyi14}.

\begin{table*}[]
\caption{Scissors resonance parameters for $^{243}$Pu and sum-rule estimates.}
\begin{threeparttable}
    \centering
\begin{tabular}{c|cccc|cccc|cc|cc}
\hline
\hline
Deformation&\multicolumn{4}{c}{Lower resonance}
&\multicolumn{4}{|c}{Upper resonance}&\multicolumn{2}{|c}{Total}
&\multicolumn{2}{|c}{Sum rule}\\
\hline
$\delta$&$\omega_{\rm SR,1}$&$\sigma_{\rm SR,1}$&$\Gamma_{\rm SR,1}$
& $B_{\rm SR,1}$ &$\omega_{\rm SR,2}$&$\sigma_{\rm SR,2}$&$\Gamma_{\rm SR,2}$
& $B_{\rm SR,2}$ &$\omega_{\rm SR}$& $B_{\rm SR}$ &      $\omega_{\rm SR}$
& $B_{\rm SR}$ \\
    &    (MeV) &(mb)    & (MeV)  &($\mu_N^2$)  &      (MeV)&(mb)     
    & (MeV)   &($\mu_N^2$)  &      (MeV)   &($\mu_N^2$)  &   (MeV)
    & ($\mu_N^2$) \\
\hline
0.27\tnote{a}&  1.99(4) & 0.45(6)& 0.60(8)&   4.8(9)    &  2.81(5)  
& 0.51(8)  & 0.83(14) & 5.3(12)     &  2.42(5)     & 10.1(15)    &   2.3  
& 10.6      \\
\hline
\hline
\end{tabular}
\begin{tablenotes}
\item[a] Average of calculations using the ground state deformation 
parameter $\beta_2$ from Refs.~\cite{RIPL3,goriely2009}.
\end{tablenotes}
\end{threeparttable}
\\
\label{tab:strengths}
\end{table*}

Recent microscopic calculations revealed that the SR contains two
modes~\cite{Balbutsev2015}, which could explain the splitting seen
experimentally in the actinides. The traditional mode consists of protons
oscillating against neutrons and a new ``nuclear spin scissors mode"
consisting of oscillations of nucleons with the spin projection ``up"
against nucleons with the spin projection ``down". The latest calculations
include spin degrees of freedom and pairing, and show good agreement with
experimental data for rare earth nuclei. 

Calculations using the sum-rule approach~\cite{lipparini1989}, were made
to predict both the centroid $\omega_{\rm SR}$ and strength $B_{\rm SR}$
of the scissors mode. The description of Enders {\em et al.}~\cite{enders2005}
was followed. The ground-state moment of inertia was replaced by the rigid-body
moment of inertia. The sum rule for the 
quasi-continuum was recently presented~\cite{gsf2014}, and a detailed
description of the formulae and parameters used for $^{238}$Np is given
in Ref.~\cite{tornyi14}. The same approach has been applied here. The
inversely and linearly energy-weighted sum rules are given by~\cite{gsf2014}:

\begin{eqnarray}
S_{+1}&=&\frac{3}{2\pi}\Theta_{\rm rigid}\delta^2\omega_D^2
\left(\frac{Z}{A}\right)^2\xi ~\left[ \mu^2_N {\rm MeV} \right], \\
S_{-1}&=&\frac{3}{16\pi}\Theta_{\rm rigid}\left(\frac{2Z}{A}
\right)^2 ~\left[ \mu^2_N {\rm MeV}^{-1} \right],
\end{eqnarray}
where $\Theta_{\rm rigid}$ is the rigid moment of inertia, $\xi$ the reduction
factor, and $\omega_D$ the iso-vector giant dipole resonance IVGDR frequency.
The nuclear quadrupole deformation $\delta=0.27$ is obtained using the ground
state deformation parameter $\beta_2$. To lowest order the two quantities are
proportional: $\delta \approx \beta_2 \sqrt{45/(16\pi)}$. The ground state
deformation taken is the average of the RIPL tabulated value~\cite{RIPL3} for
$^{242}$Pu and $^{244}$Pu ($\beta_2=0.29$) and from a microscopic
calculation~\cite{goriely2009} ($\beta_2=0.28$).
The two sum rules can now be utilized to extract the SR centroid
$\omega_{\rm SR}$ and strength $B_{\rm SR}$:
\begin{eqnarray}
\omega_{\rm SR}&=& \sqrt{S_{+1}/S_{-1}} \nonumber\\  
B_{\rm SR}&=& \sqrt{S_{+1}S_{-1}}. 
          \label{eq:b}
\end{eqnarray}

The two last columns of Table~\ref{tab:strengths} show the predicted 
$\omega_{\rm SR}$ and $B_{\rm SR}$ 
from the sum-rule estimates. Both values are in very good agreement 
with our measurements.

\section{Hauser Feshbach calculations of the $^{242}$Pu($n,\gamma$) 
cross section with TALYS}
The $\gamma$SF is important for the description of the $\gamma$ emission
channel in nuclear reactions and is one of the main inputs for cross section
calculations using a statistical framework. Calculations made with the TALYS
code~\cite{koning2008,koning12} for $^{238}$Np have shown excellent agreement
with measured data and that the SR can have an impact on the cross section
(maximum of $\approx25\%$ for a 1~MeV incident neutron~\cite{tornyi14}). 

\begin{figure}[t]
  \begin{center}
 \includegraphics[clip,width=\columnwidth]{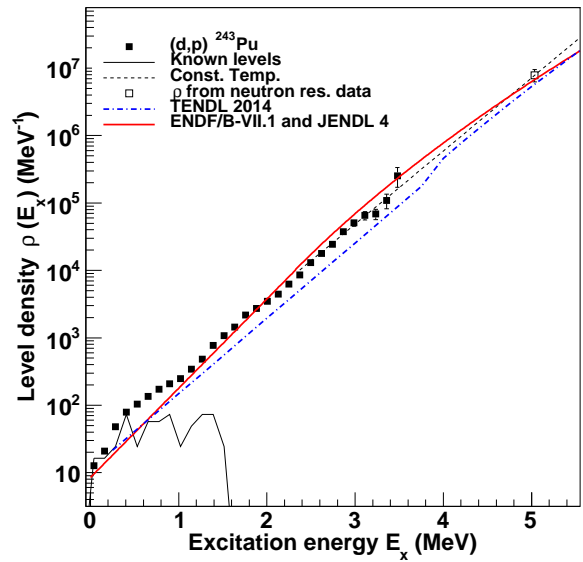}
  \caption{(Color online) Measured level density (black filled squares)
  compared to the level density used in ENDF/B-VII.1~\cite{ENDF/B-VII.1}
  and JENDL-4.0~\cite{JENDL-4.0} (red continuous curve), and
  TENDL2014~\cite{koning2008} (blue dotted-dashed curve) calculations.
  The measured level density was normalized to the level density of known
  levels (black line) and to the level density extracted from known neutron
  resonance spacings $D_0$~\cite{mughabghab2006} (empty square).}
  \label{fig:nldcomparison}
  \end{center}
  \end{figure}

Unfortunately, there are no measured data for the $^{242}$Pu$(n,\gamma)$
reaction for neutron energies above 100~keV. A comparison of the
$^{242}$Pu$(n,\gamma)$ cross section with the
ENDF/B-VII.1~\cite{ENDF/B-VII.1}, JENDL-4.0~\cite{JENDL-4.0}
and TENDL2014~\cite{TENDL2014} was done. ENDF/B-VII.1, and JENDL-4.0 are
using the same models and input parameters to calculate the level density
and the $\gamma$SF. Figure~\ref{fig:nldcomparison} shows the level densities
used by ENDF/B-VII.1, JENDL-4.0, and TENDL2014 to calculate cross sections,
as well as the level density measured in the present work. TENDL2014,
ENDF/B-VII.1, and JENDL-4.0 are within a factor of 2 from the measured level
density in the present work. 

\begin{figure}[t]
  \begin{center}
 \includegraphics[clip,width=\columnwidth]{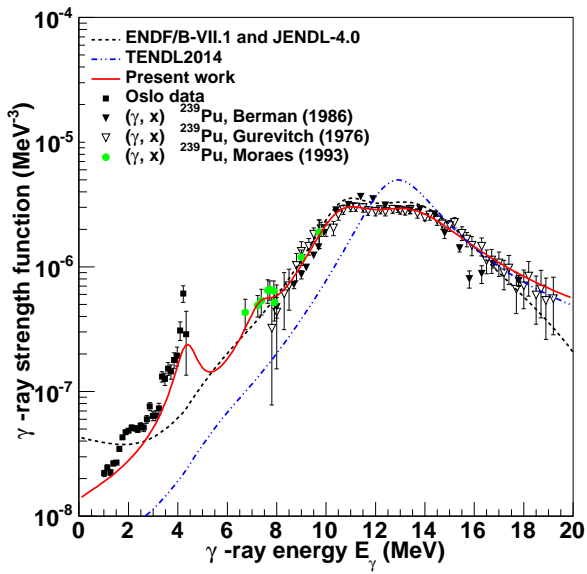}
  \caption{(Color online) Comparison between the $\gamma$SF extracted in
  the present work (red curve) from measured data (black filled squares)
  and the one used in ENDF/B-VII.1 and JENDL-4.0 (dashed black curve),
  TENDL2014 (blue dashed-dotted curve). The ($\gamma$, x) data (black
  filled triangles, empty triangles and green dots) are taken respectively
  from Berman {\em et al.}~\cite{berman1986},
  Gurevitch {\em et al.}~\cite{gurevitch1976} and
  Moraes {\em et al.}~\cite{Moraes1989}. }
  \label{fig:ySFcomparison}
  \end{center}
  \end{figure}

Figure~\ref{fig:ySFcomparison} shows the $\gamma$SF used in ENDF/B-VII.1,
JENDL-4.0 and TENDL2014 to calculate cross sections, as well as the one
measured in the present work. ENDF/B-VII.1 and JENDL-4.0 reproduce
correctly the measured ($\gamma$, x) 
data~\cite{berman1986, gurevitch1976, Moraes1989}. 
The low
energy region does not correspond well to the data measured in the present
work. TENDL2014 does not reproduce published ($\gamma$, x) data.

To calculate the $^{242}$Pu$(n,\gamma)$ cross section, the observed
level density and $\gamma$SF (data from Figs.~\ref{fig:rhotot} and
\ref{fig:GDR}, respectively) have been used as input parameters in TALYS.
The average resonance spacing $D_0$ and the average radiative width
$\langle\Gamma_\gamma\rangle$, from Ref.~\cite{mughabghab2006}, 
are reproduced by the TALYS calculation.
The neutron optical potential is taken from Ref.~\cite{actinideOMP}.

\begin{figure}[b]
 \begin{center}
\includegraphics[clip,width=\columnwidth]{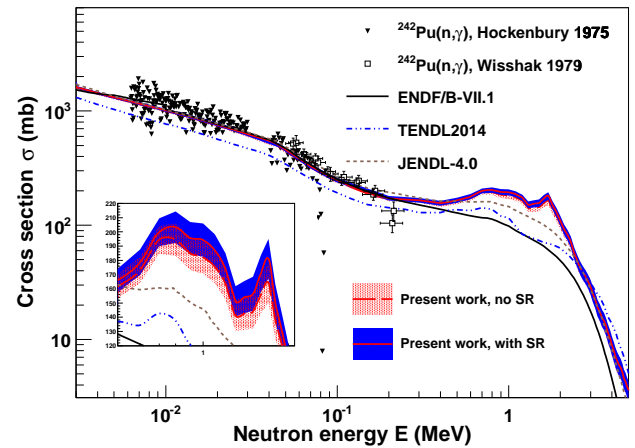}
 \caption{(Color online) Calculated $^{242}$Pu$(n, \gamma)$ cross section using
 level density and $\gamma$SF parameters obtained in the present work,
 including the $M1$ scissors mode (continuous red curve with blue error-band)
 and without it (dashed red curve with red dots error-band). A zoom in the
 energy region 0.5 to 2~MeV, where the impact of the SR is the most important,
 is shown in the inset. The predictions are compared at low energy with
 measured data from Hockenbury {\em et al.}~\cite{Hockenbury1975}
 (black triangles) Wisshak and K{\"a}ppeler~\cite{Wisshak79} (empty squares), 
 and evaluations from ENDF/B-VII.1 (black curve),
 JENDL-4.0 (dashed grey curve), TENDL2014 (blue dotted-dashed curve).}
 \label{fig:cross}
 \end{center}
 \end{figure}

Figure~\ref{fig:cross} shows the results of the cross-section calculations
using the TALYS code with the SR (continuous red curve with blue error-band)
and without (dashed red curve with red dots error-band), the
ENDF/B-VII.1 (black curve), JENDL-4.0 (brown curve) and
TENDL2014 (blue curve) evaluations. The error band is generated by taking into
account the uncertainty in the two pygmy resonances labeled pygmy1 and pygmy2
in Fig.~\ref{fig:GDR} and the average radiative width
$\langle\Gamma_\gamma\rangle$. Including the SR in the calculation leads to some
variations in the cross section (up to $\approx$10$\%$ at 1.7~MeV). This is
smaller in comparison to the recent measurement on $^{238}$Np~\cite{tornyi14}
with a comparable SR strength.

Below 200~keV, the data libraries and our calculations agree with the direct 
measurement from Hockenbury {\em et al.}~\cite{Hockenbury1975} (black triangles) 
and the cross section data from Wisshak {\em et al.}~\cite{Wisshak79}, obtained 
as a ratio to the $^{197}$Au(n,$\gamma$) cross section. The open squares are 
extracted using the   $^{197}$Au(n,$\gamma$) cross section from the IRDFF-1.05 
database~\cite{IRDFF}.
At higher energies, large discrepancies are observed between 
the different libraries and our
calculation as can be expected due to the discrepancies in the level densities
and $\gamma$SF and the lack of directly measured experimental data at higher
neutron energies. Surprisingly the ENDF/B-VII.1 and JENDL-4.0 cross sections
do not match despite using the same level density and $\gamma$SF.
The ENDF/B-VII.1 cross section was re-normalized to an integral measurement
over a broad fast spectrum~\cite{Dr77}. Direct measurements of
the $^{242}$Pu(n,$\gamma$)
cross section are planned at the n$\_$TOF facility at CERN and should help
solve the discrepancy. 

\section{Conclusions and future work}
\label{sec:con}

The level density and $\gamma$SF of $^{243}$Pu have been measured in the
quasi-continuum using the Oslo method. The level density follows closely a
constant-temperature level density formula as seen in recent investigations
of other actinides using the same method~\cite{nld2013,gsf2014,tornyi14}.
The $\gamma$SF displays a double-humped resonance in the
$E_{\gamma} = 1 - 3.5$~MeV region, interpreted as the scissors resonance.
Its energy centroid and total strength are very well described by the
sum-rule estimate assuming a rigid-body moment of inertia. 

The observed level density and $\gamma$SF have been used as inputs in
Hauser-Feshbach calculations implemented in the TALYS code. Large
discrepancies with the ENDF, JENDL and TENDL databases raise the need
for a direct measurement of the  $^{242}$Pu(n,$\gamma$) cross section.

A $^{244}$Pu target has recently been made at Lawrence Livermore
National Laboratory. Experiments at the Oslo Cyclotron Laboratory
using the ($^3$He,$\alpha$) and/or $(p,d)$ entrance channels are of
interest to compare with the results presented here and study the effect
of spin population of the compound nucleus on the extracted statistical
properties of the nucleus.

\acknowledgements

We would like to thank J.C.~M{\"{u}}ller, E.A.~Olsen, A.~Semchenkov and
J.C.~Wikne at the Oslo Cyclotron Laboratory for providing the stable and
high-quality deuterium beam during the experiment. 

This work was performed under the auspices of the U.S. Department of Energy
by Lawrence Livermore National Laboratory in part under Contract W-7405-Eng-48
and in part under Contract DE-AC52-07NA27344, the U.S. Department of Energy by
Lawrence Berkeley National Laboratory under Contract No. DE-AC02-05CH11231, the
University of California Office of the President Laboratory Fees Research
Program under Award No. 12-LR-238745. We also wish to acknowledge support 
from the Peder Sather Center for Advanced Study at the University of 
California, Berkeley. A.C.L. acknowledges financial support from the 
ERC-STG-2014 under grant agreement no.~637686. Financial support from the
Norvegian Research Council grant no.~205528 (A.C.L.), grant no.~222287 (G.M.T.)
and grant no.~210007 (F.G. and S.S.) are gratefully acknowledged.
M.W. acknowledges support from the National
Research Foundation of South Africa under grant no.~92789.

\vfill
\end{document}